\documentclass[prb,preprint,superscriptaddress]{revtex4}
\usepackage{amsmath}
\usepackage{amssymb}
\usepackage{amsbsy}
\usepackage{graphicx}
\newcommand{\unit}[2]%
{\mbox{\ensuremath{#1}}\mbox{\,\ensuremath{\mathrm{#2}}}}
\newcommand{\Tc}{\ensuremath{T_\mathrm{c}}}
\newcommand{\Bc}{\ensuremath{B_\mathrm{c2}}}
\newcommand{\Jc}{\ensuremath{J_\mathrm{c}}}
\newcommand{\pc}{\ensuremath{p_\mathrm{c}}}

\newcommand{\Birr}{\ensuremath{B_\mathrm{irr}}}
\newcommand{\Br}{\ensuremath{B_{\rho=0}}}

\begin{document}
\title{The influence of weak texture on the critical currents in polycrystalline MgB$_2$}
\author{Eisterer M}
\email{eisterer@ati.ac.at}
\author{Weber H W}
\affiliation{Atominstitut, Vienna University of Technology,
Stadionallee 2, 1020 Vienna}

%\pacs{74.72.Bk, 74.78.Bz, 61.80.Hg, 74.25.Qt, 74.25.Sv, 74.62.Dh}

\begin{abstract}
The current transport in polycrystalline MgB$_2$ is strongly
influenced by the intrinsic anisotropy of this superconductor.
Untextured bulks and wires are macroscopically isotropic, but the
grains retain their anisotropic properties and the field
dependence of the critical currents is much stronger than in
isotropic superconductors. Weakly or partially textured tapes are
macroscopically anisotropic, but the anisotropy of the zero
resistivity (or irreversibility) field is smaller than the
intrinsic upper critical field anisotropy, $\gamma$. The
\Jc-anisotropy is field and temperature dependent and can be much
larger than $\gamma$. The most suitable parameter for the
quantification of the macroscopic anisotropy is, therefore, the
anisotropy of the zero resistivity field. It is difficult to
distinguish between a higher degree of texture at a lower
intrinsic anisotropy and a weaker texture at higher anisotropy and
hardly possible based on the field dependence of the critical
current anisotropy alone. The knowledge of the upper critical
field is crucial and angular resolved measurements of either the
critical currents or, better, the resistive in-field transitions
are favorable for this purpose.

\end{abstract}
\maketitle

\section{Introduction}
A considerable anisotropy of the critical current is found in
\emph{ex-situ} \cite{Gra02,Kit05b,Lez05b,Bei06,Lez06b,Ser08} and
\emph{in-situ} \cite{Kov05,Kov08} MgB$_2$ tapes. It is favored by
a large aspect ratio of the tape, and is a consequence of the
upper critical field anisotropy, $\gamma$, in magnesium diboride
\cite{Eis07rev,Lya02,Bud02,Wel03b,Ang02}, if partial or weak
texture is induced by the preparation technique. If the grain
orientation is totally random (no texture), the conductor is
macroscopically isotropic, since no preferred orientation exists.
However, also in that case the intrinsic anisotropy of MgB$_2$ has
severe consequences on the properties of polycrystalline
materials. The critical current density, \Jc, becomes zero well
below the upper critical field, $B_{c2}$, which in turn leads to a
much stronger field dependence of the critical current than in
isotropic materials \cite{Eis03}. The Kramer plot is no longer
linear near the irreversibility field and the peak of the volume
pinning force shifts to significantly lower fields \cite{Eis08}.
Large critical current densities of the order of $5\times
10^8$\,Am$^{-2}$ can be expected only up to \cite{Eis05}
$B_{c2}^c=B_{c2}^{ab}/\gamma$, since only highly percolative
currents flow on a macroscopic length scale at higher fields.
$B_{c2}^{ab}$ and $B_{c2}^c$ denote the upper critical field, when
the field is applied parallel and perpendicular to the boron
planes, respectively. $B_{c2}^{ab}$ is around 14\,T at 0\,K in
clean MgB$_2$ and determines the upper critical field in
untextured polycrystalline samples \cite{Eis05}, if defined by a
typical onset criterion (e.g. our definition in
Sec.~\ref{SecRes}). $B_{c2}^c$ is not easily accessible in
polycrystalline MgB$_2$ and only about 3\,T at 0\,K in the clean
limit \cite{Eis07rev,Lya02,Bud02,Wel03b}, which strongly restricts
the field range for applications. Fortunately, $B_{c2}^{ab}$ (and
the upper critical field in polycrystalline samples) can be
enhanced by impurity scattering \cite{Gur03,Eis07,Put05}, which
also reduces $\gamma$ \cite{Kru07,Eis07rev}. Both changes enhance
$B_{c2}^c$ at low temperatures, which is good for applications. Impurity scattering is less
effective in improving the properties at high temperatures, since
it decreases the transition temperature \cite{Put07} and this change offsets
the other beneficial effects. Texturing the material
might, therefore, be an interesting alternative for optimizing the
material for application at higher temperatures and fields. It
increases the in-field critical currents for one orientation at
the expense of decreasing them in the other main orientation
\cite{Eis09d}. Texture is therefore only beneficial, if the tape
is operated in or near the favorable orientation, which generally
complicates the design of devices, but is certainly feasible in
superconducting magnets.

Only weak texture was achieved in MgB$_2$ tapes so far, with the
c-axis oriented preferentially perpendicular to the tape. The
average misalignment angle of the $c$-axis from the tape normal
was found to be rather high \cite{Lez06b,Hae09} (above 20$^\circ$)
compared to highly textured materials (a few degrees only).

Weak texture is a unique property of some MgB$_2$ tapes, since the
strongly anisotropic cuprate superconductors have to be highly
textured for high intergranular currents \cite{Fel08,Dur09}
leading to a pronounced \Jc-anisotropy. NbTi and Nb$_3$Sn on the
other hand are isotropic superconductors, thus (weak) texture is
not expected to influence the properties of wires or tapes. In
this report the influence of weak texture on the macroscopic
current transport in MgB$_2$ will be discussed and the resulting
behavior compared to the limiting cases of zero or perfect
texture.

\section{The model}

A general model for nonlinear transport in heterogeneous media was
presented in Ref.~\onlinecite{Eis03}. The critical current was
obtained by integrating the percolation cross section,
$\sigma_\mathrm{p}\propto (p(J)-p_\mathrm{c})^t$, over all
possible local currents $J$. $p(J)$ was defined as the fraction of
grains with a $J_\mathrm{c}^\mathrm{local}$ above $J$. The
percolation threshold, \pc, depicts the minimum fraction of
superconducting grains for a continuous path through the
superconductor. The percolation cross section reduces to zero at
\pc, which is expected to be around 0.25 in typical
polycrystalline MgB$_2$ materials \cite{Eis03,Gri06,Yam07,Eis09}.
This value will be used in the following. The transport exponent,
$t$, is expected to be 1.76 \cite{Deu83}, which will be assumed in
the following, although higher values were also observed
experimentally \cite{Vio05,Eis09}. Furthermore, a model for the
field dependence of the local critical currents is needed for
calculating $p(J)$. Since grain boundary pinning is dominant in
the majority of MgB$_2$ conductors \cite{Eis03,Mar07,Eis07rev},
$J_\mathrm{c}^\mathrm{local}\propto (1-b)^2/\sqrt{B}$ was assumed
\cite{Dew74}, but contributions of point pins were also reported
\cite{Pal05,Che06b}. $b$ denotes the magnetic field normalized by
the (angular dependent) upper critical field of the grain.
Finally, the anisotropic scaling approach \cite{Bla92} was used to
model the angular dependence of the local critical currents. Note
that we already had to make four approximations/assumptions (\pc,
$t$, pinning model, anisotropic scaling), which influence the
field dependence of \Jc\ independently of the degree of texture.

Texture is a modification of the distribution function of the
grains' orientation. For uniaxial texture and the present geometry
(preferred $c$-axis orientation perpendicular to the tape), the
density of grains with a certain orientation only depends on the
angle $\alpha$ between the tape normal and the $c$-axis of the
grains. The corresponding distribution function is a priori
unknown. A Gaussian-like exponential function
\begin{equation}
f_\mathrm{a}(\alpha) \propto
\exp(-\frac{\alpha^2}{2\alpha_\mathrm{t}^2}) \label{Gauss}
\end{equation}
was chosen, which was confirmed by high energy synchrotron x-ray
diffraction measurements\cite{Hae09}. Since $\alpha$ is restricted
to $[0,90^\circ]$, the density distribution function has to be
renormalized numerically \cite{Eis09d} and is formally not a true
Gaussian distribution. This implies that $\alpha_\textrm{t}$ is no
longer the standard deviation of the distribution or an average
misalignment angle, which are restricted to angles below
90$^\circ$, but only a parameter which characterizes the decrease
of the density distribution function between  0 and 90$^\circ$
and, therefore, the degree of texture. $\alpha_\mathrm{t}$
converges to zero in the limit of perfect texture and diverges in
untextured materials ($f_\mathrm{a}$ becomes constant between 0
and 90$^\circ$ for $\alpha_\mathrm{t}\rightarrow \infty$, uniform
distribution).

Only a few quantitative data about texture in MgB$_2$ tapes exist
in the literature, varying between 20$^\circ$ and 30$^\circ$
\cite{Lez06b, Hae09}, but the model is not restricted to this
angular range. For $\alpha_\mathrm{t}\rightarrow 0$ and $\infty$,
the results of the model merge the predictions of the anisotropic
scaling approach \cite{Bla92} and the percolation model for
untextured MgB$_2$ \cite{Eis03}, respectively.

For the calculation of the angular dependence of \Jc, \Bc\ and
\Br, the density distribution function with respect to the
direction of the applied magnetic field is needed, which was
derived from the distribution function Equ.~\ref{Gauss} in
Ref.~\onlinecite{Eis09d}. All following calculations are based on
this density distribution function.

\section{Results and Discussion \label{SecRes}}

\begin{figure} \centering \includegraphics[clip,width=0.45\textwidth]{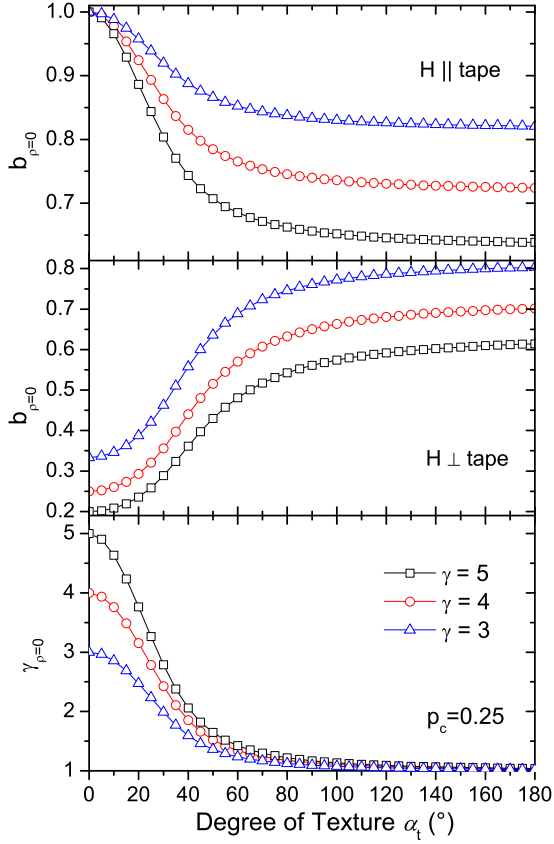}
\caption{The zero resistivity field in parallel orientation (top)
increases with the degree of texture (decreasing
$\alpha_\mathrm{t}$), but decreases in perpendicular orientation
(middle). \Br\ was normalized by $B_\mathrm{c2}^{ab}$
($b_\mathrm{\rho=0}:=B_\mathrm{\rho=0}/B_\mathrm{c2}^{ab}$). The
anisotropy of \Br\ varies between the intrinsic value at
$\alpha_\mathrm{t}=0$ and 1 in the untextured material (bottom).
It changes strongest between $10^\circ$ and $70^\circ$.}
\label{Figbirr}
\end{figure}

At first we focus on the anisotropy of the zero resistivity
(irreversibility) field \Br. This quantity is less model dependent
than \Jc. It is independent of the pinning mechanism, the
transport exponent and anisotropic scaling. Only the angular
dependence of the upper critical field has to be known. Good
agreement with the theoretical prediction of the anisotropic
Ginzburg Landau theory \cite{Til65},
\begin{equation}
\Bc(\theta)=B_\mathrm{c2}^{ab}/\sqrt{\gamma^2\cos^2(\theta)+\sin^2(\theta)},
\label{equGL}
\end{equation}
was demonstrated \cite{Lya02,Ang02,Pri03b,Ryd04}. It was shown in
Ref.~\onlinecite{Eis05}, that \Br\ in untextured materials is
given by
\begin{equation}
B_{\rho=0}=\frac{B_\mathrm{c2}^{ab}}{\sqrt{(\gamma^{2}-1)p_\mathrm{c}^{2}+1}}.
\label{equrho0}
\end{equation}
While $B_\mathrm{c2}^{ab}$ can be easily assessed experimentally,
\pc\ \cite{Eis09} and $\gamma$ \cite{Bud02} are difficult to
determine. By fixing \pc\ to a realistic value, $\gamma$ can be
estimated from the difference between\cite{Kim08} \Br\ and
$B_\mathrm{c2}^{ab}$, or, alternatively, by
estimating $\gamma$ from \Tc, \pc\ can be obtained
\cite{Eis07rev,Eis09}. However, any uncertainty in one quantity
induces a similar error in the other one; furthermore
Equ.~\ref{equrho0} was derived for a perfectly homogeneous
material and neglecting the influence of thermal fluctuations.
Both effects increase the difference between \Br\ and \Bc. Thermal
fluctuations, which result in a difference between the
\emph{intrinsic} \Bc\ and \Birr\ in each grain, generally increase
with \Bc\ (the importance of fluctuations is quantified by the Ginzburg number \cite{Bra95} $Gi\propto B_\textrm{c2}^3$) and $T$. A successful description of the resistive
transition by Equ.~\ref{equrho0} thus requires a homogeneous
material (narrow superconducting zero field transition), a
moderate upper critical field, and temperatures well below \Tc.

Texture enhances \Br\ (compared to the prediction of
Equ.~\ref{equrho0} for untextured MgB$_2$), when the field is
oriented parallel to the tape ($B_\mathrm{\rho=0}^\parallel$, top
panel in Fig.~\ref{Figbirr}), but reduces \Br\ in perpendicular
orientation ($B_\mathrm{\rho=0}^\perp$, middle panel) leading to
an anisotropy of $B_\mathrm{\rho=0}$,
$\gamma_{\rho=0}:=B_\mathrm{\rho=0}^\parallel
/B_\mathrm{\rho=0}^\perp$ (bottom panel). The \emph{intrinsic}
anisotropy is obtained at $\alpha_\mathrm{t}=0^\circ$ (perfect
texture), the anisotropy is very small at
$\alpha_\mathrm{t}=180^\circ$ (1.02, 1.03, and 1.04 for
$\gamma=3$, 4, and 5, respectively).

\begin{figure} \centering \includegraphics[clip,width=0.45\textwidth]{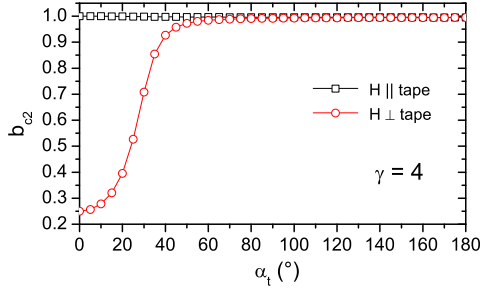}
\caption{The upper critical field is less sensitive to texture
than \Br\ (cf. Fig.~\ref{Figbirr}). \Bc\ was normalized by
$B_\mathrm{c2}^{ab}$
($b_\mathrm{c2}:=B_\mathrm{c2}/B_\mathrm{c2}^{ab}$) and refers to
the onset of the superconducting transition.} \label{FigBc2}
\end{figure}

The upper critical field is less sensitive to texture than \Br, as
demonstrated for $\gamma=4$ in Fig.~\ref{FigBc2}. The data roughly
correspond to a 90\% resistive criterion, since a fraction of
2.5\% superconducting grains was assumed for the calculations. If,
say, 25\% superconducting grains were needed for zero resistivity
($\pc=0.25$), around 2.5\% should be enough to lower the
resistivity by 10\%. This definition (2.5\% superconducting
grains) will be used for the (angular dependent) upper critical
field, \Bc, in the following. (Henceforth $B_\mathrm{c2}(\theta_\mathrm{t})$ always denotes this macroscopic upper critical and not the intrinsic $B_\mathrm{c2}(\theta)$
(Equ.~\ref{equGL}) within the grains.) The upper critical field is
nearly independent of texture, if the field is applied parallel to
the tape. This is why the onset of the transition always
corresponds to $B_\mathrm{c2}^{ab}$ in untextured materials. A
significant anisotropy of the upper critical field is observed
only for $\alpha_\mathrm{t}$ below about $45^\circ$. The
anisotropy of the upper critical field is only 1.4, while
$\gamma_{\rho=0}$ is 2.4 at $\alpha_\mathrm{t}=30^\circ$.

The upper critical field and its anisotropy are much more
sensitive to the chosen criterion than \Br\ and $\gamma_{\rho=0}$.
Therefore, and because of the field and temperature dependence of
the \Jc-anisotropy (see below),  $\gamma_{\rho=0}$ is the most
useful parameter for the quantification of the macroscopic
anisotropy.

\begin{figure} \centering \includegraphics[clip,width=0.45\textwidth]{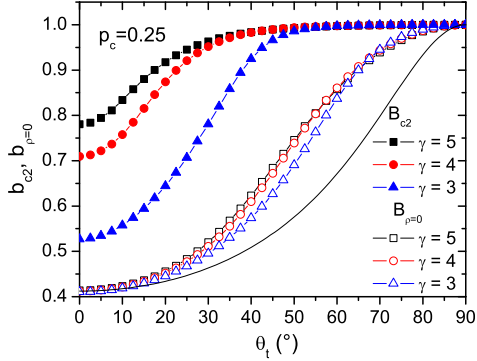}
\caption{Angular dependence of the upper critical field and the
zero resistivity field. $\theta_\mathrm{t}$ denotes the angle
between the tape normal and the applied magnetic field. \Bc\ and
\Br\ were normalized by their values at 90$^\circ$. The
\Br-anisotropy was fixed to 2.4 which corresponds to
$\alpha_\mathrm{t}=34.5$, 30, and 20.8$^\circ$ for $\gamma=5$, 4,
and 3, respectively. Neither \Bc\ nor \Br\ obey the angular
dependence predicted by anisotropic Ginzburg Landau theory (solid
line).} \label{FigBtheta}
\end{figure}

The same $\gamma_{\rho=0}$ (2.4) can be obtained with $\gamma=5$
and $\alpha_\mathrm{t}=34.5^\circ$ or with $\gamma=3$ and
$\alpha_\mathrm{t}=20.8^\circ$. The corresponding angular
dependencies of \Br\ are presented in Fig.~\ref{FigBtheta} (open
symbols). They are similar to each other and cannot be described
by the angular dependence predicted by anisotropic Ginzburg Landau
theory (Equ.~\ref{equGL}, line graph) by assuming
$\gamma_{\rho=0}$ instead of $\gamma$. The deviation from this
theory is even more pronounced for the onset of the transition
($B_\mathrm{c2}$, solid symbols) and the anisotropy of the upper
critical field is quite different at the same \Br-anisotropy. It
seems therefore possible to estimate the three unknown parameters
($\gamma$, $\alpha_\mathrm{t}$, and \pc) from the angular
dependence of \Br\ and \Bc\ and from the ratio of these quantities
(not visible in Fig.~\ref{FigBtheta}, because the data were
normalized by the values in parallel orientation).

\begin{figure} \centering \includegraphics[clip,width=0.45\textwidth]{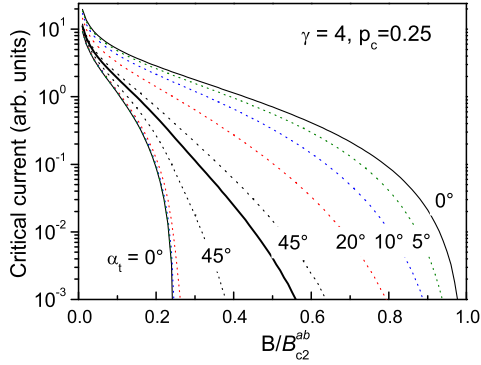}
\caption{Influence of texture on the critical currents. The thick
solid line refers to untextured material
($\alpha_\mathrm{t}\rightarrow\infty$), dotted lines above and
below to the parallel and perpendicular field orientation,
respectively.} \label{FigJc}
\end{figure}

The increase of \Br\ in parallel orientation as well as the
decrease in perpendicular orientation (compared to untextured
materials) lead to a weaker and stronger field dependence of the
critical currents, respectively (Fig.~\ref{FigJc}). While \Jc\
converges to the limiting behaviour of perfect texture
($\alpha_\mathrm{t}=0^\circ$) rather quickly in the perpendicular
orientation, it remains significantly below its limit in the
parallel orientation, even for a high degree of texture
($\alpha_\mathrm{t}=5^\circ$). However, for a typical texture of
$\alpha_\mathrm{t}=20$--30$^\circ$ \cite{Lez06b,Hae09}, a quite
significant enhancement beyond untextured materials can be
expected at $B=B_\mathrm{c2}^{ab}/2$ in the parallel orientation.

\begin{figure} \centering \includegraphics[clip,width=0.45\textwidth]{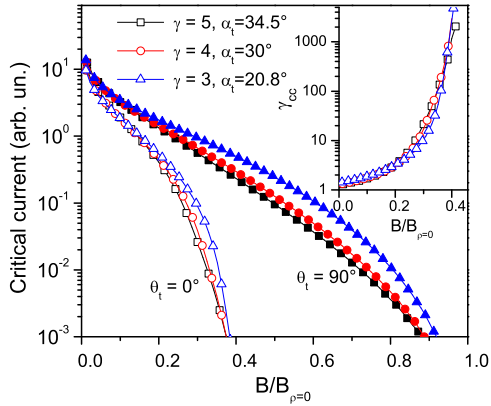}
\caption{Critical currents in materials with the same anisotropy
of \Br\ ($\gamma_{\rho=0}=2.4$), but different degree of texture
and intrinsic anisotropy $\gamma$. The magnetic field is
normalized by the zero resistivity field in parallel field
configuration. The inset presents the anisotropy of the critical
currents $\gamma_\mathrm{cc}$.} \label{FigJcgamma}
\end{figure}

Next we consider the influence of different combinations of
$\gamma$ and $\alpha_\mathrm{t}$, which lead to the same
$\gamma_{\rho=0}=2.4$, on the field dependence of \Jc\
(Fig.~\ref{FigJcgamma}). Although the results are slightly
changed, the differences are not sufficient to derive these two
parameters from \Jc$(B)$ alone, since they are also influenced by
the usually unknown percolation threshold, the transport exponent,
the pinning model and the angular dependence of
$J_\mathrm{c}^\mathrm{local}$. Moreover \Jc$(B)$ is usually not
available over four orders of magnitude. Note on the other hand
that $B_\mathrm{\rho=0}^{\parallel}/B_\mathrm{c2}^{ab}$ is
different (0.7, 0.8, and 0.91 for
$\gamma$/$\alpha_\mathrm{t}=5/34.5^\circ, 4/30^\circ$, and
3/20.8$^\circ$, respectively), which cannot be seen in
Fig.~\ref{FigJcgamma} due to the normalization. The knowledge of
the upper critical field is therefore vital for the distinction
between $\gamma$ and $\alpha_\mathrm{t}$. Unfortunately, the metal
sheath of tapes usually masks the onset of the transition
\cite{Eis05} and must be removed for the determination of \Bc.

The anisotropy of the critical current is strongly field dependent
(inset in Fig.~\ref{FigJcgamma}) and diverges at
$B_\mathrm{\rho=0}^\perp$ (\Br\ in orthogonal orientation). At
fixed reduced field, $B/\Br$, it is not very sensitive to changes
of the parameters, as long as the \Br-anisotropy is kept constant
and is, therefore, also not appropriate for a (simultaneous)
determination of $\gamma$ and $\alpha_\mathrm{t}$. Since the field
dependence of the critical current anisotropy scales with the
upper critical field, the critical current anisotropy is also
temperature dependent, because the reduced magnetic field
increases with temperature at fixed magnetic field (\Bc\
decreases).

\begin{figure} \centering \includegraphics[clip,width=0.45\textwidth]{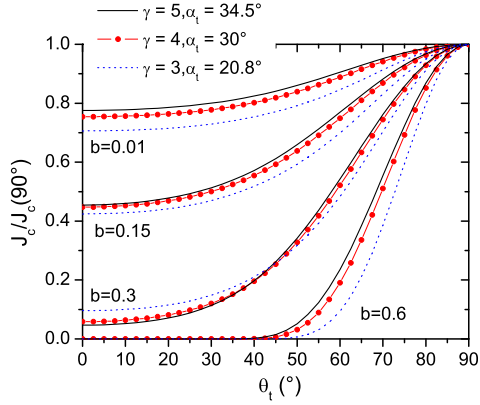}
\caption{The angular dependence of \Jc\ is altered by different
combinations of $\gamma$ and $\alpha_\mathrm{t}$, which lead to
the same anisotropy of the zero resistivity field, \Br\
($\gamma_{\rho=0}=2.4$). The reduced fields, $b$, were normalized
to \Br\ in parallel orientation.} \label{FigangJc}
\end{figure}

The angular dependence of \Jc\ is presented in
Fig.~\ref{FigangJc}. The same combinations of $\gamma$ and
$\alpha_\mathrm{t}$ as for the field dependence of \Jc\ were used.
Data at the same reduced field, $b=B/B_\mathrm{\rho=0}^\parallel$
are compared. The differences between $\gamma=4$ and $\gamma=5$
are small, but the curve for $\gamma=3$ differs more
significantly, which favors the analysis of texture from angular
resolved measurements of \Jc. However, the higher degree of
modeling (pinning mechanism, percolation cross section,
anisotropic scaling) renders this procedure less reliable than
measurements of the resistive transitions (without sheath) at
various angles.

The data in Fig.~\ref{FigangJc} can be fitted to the behaviour
predicted by the anisotropic scaling approach \cite{Bla92} for
perfectly textured materials, with \Bc\ and $\gamma$ as free
parameters (not shown in Fig.~\ref{FigangJc}). However, the
obtained parameters do not have any physical meaning, because they
change with field and strongly underestimate the real \Br\
($<\Bc$) and $\gamma_{\rho=0}$ ($<\gamma$).

\section{Conclusions}
The influence of homogeneous weak texture on the macroscopic
currents in MgB$_2$ was investigated on the basis of a percolation
model for nonlinear transport in heterogeneous media. While any
texture leads to an anisotropy of the zero resistivity field, the
occurrence of a macroscopic upper critical field anisotropy (onset
of the superconducting transition) indicates a significant degree
of texture. Although the calculations were based on the
anisotropic scaling approach for the intrinsic properties, the
angular dependence of the macroscopic quantities (i.e., critical
current, zero resistivity field, and upper critical field) do not
obey the predictions of this general scaling in weakly textured
materials, even if an effective instead of the intrinsic
anisotropy is assumed.

It is hardly possible to derive the degree of texture and the
intrinsic anisotropy from the field dependence of the critical
current, in particular, if the upper critical field is unknown.
Angular resolved measurements of the resistive transition turn out
to be more appropriate for this purpose, but usually require the
removal of the sheath.

The most suitable parameter for the quantification of the
macroscopic anisotropy is the anisotropy of the zero resistivity
field.

%\begin{acknowledgments}

%\end{acknowledgments}
%\bibliographystyle{michl}
%\bibliography{ref}{}

\end{document}